\documentclass[a4paper,11pt]{article}
\pdfoutput=1 % if your are submitting a pdflatex (i.e. if you have
\usepackage{jheppub} % for details on the use of the package, please
\usepackage[T1]{fontenc} % if needed
\usepackage{slashed,color,graphicx}
\usepackage{amsmath}
\usepackage{amssymb}
\usepackage[dvipsnames]{xcolor}
\usepackage{centernot}
\usepackage{url}

\usepackage{cancel}

\newcommand{\al}{\alpha}
\newcommand{\ep}{\epsilon}

\newcommand{\de}{\delta}
\newcommand{\si}{\sigma}

\newcommand{\ol}{\overline}

\newcommand{\nl}{\nonumber\\}
\newcommand{\bea}{\begin{eqnarray}}
\newcommand{\eea}{\end{eqnarray}}

\title{\boldmath Inelastic dark matter, small scale problems, and the XENON1T excess}

\author{Seungwon Baek,}
\emailAdd{sbaek@korea.ac.kr}
\affiliation{Department of Physics, Korea University, \\
Anam-ro 145, Sungbuk-gu, Seoul 02841, Korea}

\abstract{We study a generic model in which the dark sector is composed of a Majorana dark matter $\chi_1$, its excited state $\chi_2$, both at the electroweak scale, and a light dark photon $Z'$ with $m_{Z'} \sim 10^{-4}$ eV. The light $Z'$ enhances the self-scattering elastic cross section $\chi_1 \chi_1 \to \chi_1 \chi_1$ enough to solve the small scale problems in the $N$-body simulations with the cold dark matter. The dark matter communicates with the SM via kinetic mixing parameterized by $\epsilon$. The inelastic scattering process $\chi_1 \chi_1 \to \chi_2 \chi_2$ followed by the prompt decay $\chi_2 \to \chi_1 Z'$ generates energetic $Z'$. By setting $\delta \equiv m_{\chi_2} - m_{\chi_1} \simeq 2.8$ keV and $\epsilon \sim 10^{-10}$ the excess in the electron-recoil data at the XENON1T experiment can be explained by the dark-photoelectric effect. The relic abundance of the dark matter can also be accommodated by the thermal freeze-out mechanism via the annihilation $\chi_1 \chi_1 (\chi_2 \chi_2) \to Z' Z'$ with the dark gauge coupling constant $\alpha_X \sim 10^{-3}$.}

\begin{document} 
\maketitle
\flushbottom

\section{Introduction}
\label{sec:intro}
There are clear evidences that dark matter (DM) exists in the universe,
although its particle nature is almost unknown.
Since the standard model (SM) lacks candidate for the dark matter, new physics (NP) models
are required to incorporate DM.
The standard cold dark matter (CDM) model has been very successful in predicting the large scale structure of the universe.
However, the $N$-body simulations with CDM predict cuspy density profiles, while observations of rotation curves of dwarf galaxies 
and low surface brightness galaxies point towards flat cores. Self-interacting dark matter (SIDM) with self-scattering cross sections, $\sigma_{\chi\chi}/m_{\chi} \sim 1 \,\rm{cm^2/g}$, can be
a possible solution to this problem~\cite{Tulin:2017ara}. The constraint, $\sigma_{\chi\chi}/m_\chi \lesssim 0.5\, {\rm cm^2/g}$, coming from collision of galaxy clusters~\cite{Harvey:2015hha}
can be evaded by the SIDM's velocity-dependent cross sections.
Also the non-observation of  DM in the DM-nucleon scattering experiments can be naturally explained in the inelastic DM models~\cite{Blennow:2016gde}.

In the year 2020, XENON1T collaboration has announced an excess of electron recoil near $2-3$ keV energy~\cite{Aprile:2020tmw}.
Although the result can be explained by $\beta$ decays of tritium contamination, it can also be attributed to
NP contributions such as the solar axion or anomalous neutrino magnetic moment with about 3$\sigma$ significance.
However, the latter possibilities are in strong tension with the star cooling constraints~\cite{Aprile:2020tmw}.
The excess may also be the result of the dark matter scattering with electrons inside the xenon atom.
In this case the signal can give valuable information on the nature of the dark matter.

The XENON1T excess has been considered in various inelastic DM models~\cite{Harigaya:2020ckz,Lee:2020wmh,Bramante:2020zos,An:2020tcg,Chao:2020yro,Baek:2020owl,He:2020wjs,Choudhury:2020xui,Ema:2020fit,Borah:2020jzi,Borah:2020smw,
Aboubrahim:2020iwb,He:2020sat,Dutta:2021wbn}.
In~\cite{Baek:2020owl} we studied exothermic DM scattering on electron to explain the anomaly.
We found that both scalar and fermionic DM models can accommodate the XENON1T excess.

In this paper we consider another possible realization of inelastic DM. 
As in the fermionic DM model in~\cite{Baek:2020owl} the dark sector (DS) has two Majorana DM candidates, $\chi_1, \chi_2$,
with mass spltting, $\delta = m_{\chi_2}-m_{\chi_1}$,
as well as the dark photon ($Z'$).
The current universe has only electroweak scale $\chi_1$ DM as opposed to the model 
in~\cite{Baek:2020owl}\footnote{It turns out the lifetime of $Z'$ are much longer than the age of the universe.
However, their contribution to the relic density is negligible due to $m_{Z'} \ll m_\chi$.}.
The inelastic scattering $\chi_1 \chi_1 \to \chi_2 \chi_2$ followed by the decay of the excited state $\chi_2 \to \chi_1 Z'$
produces $Z'$ with energy given by the mass splitting $\delta \equiv m_{\chi_2} - m_{\chi_1}$.
The dark photon (DP) $Z'$ whose mass is lighter than $\delta$ is absorbed in the xenon atom, ejecting an electron, which we may call the dark-photoelectric effect. 
We fix the mass splitting $\delta=2.8$ keV to fit the XENON1T excess. 

The elastic scattering $\chi_1 \chi_1 \to \chi_1 \chi_1$ can be large enough to solve the small scale problem, such as the cusp-core problem mentioned above.
Since its cross section is velocity-dependent, the constraint from the galaxy clusters can also be easily evaded.
The relic abundance of the $\chi_1$ is obtained by the thermal freeze-out mechanism.

The paper is organised as follows. In Section~\ref{sec:model} we introduce the model. In Section~\ref{sec:relic} the relic density of the dark matter is calculated.
In Section~\ref{sec:small} we show that the small scale problems of the CDM can be solved in our model.
In Section~\ref{sec:XENON1T} we consider the dark-photoelectric effect to address the excess of recoil electrons in the XENON1T.
We conclude in Section~\ref{sec:conclusion}.

\section{The model}
\label{sec:model}

The model has a dark gauge symmetry $U(1)_X$ under which all the SM particles are neutral.
A $U(1)_X$-charged, but neutral under the SM, Dirac fermion $\chi(x)$ is introduced.
For the inelastic dark matter we study the Lagrangian of the dark sector (DS) including the kinetic energy 
and kinetic mixing terms in the form~\cite{Cui:2009xq}
\begin{eqnarray}
\mathcal{L} & = & \mathcal{L}_{\rm SM} -\frac{1}{4} \hat{X}_{\mu \nu} \hat{X}^{\mu \nu} 
+ {1 \over 2} m_{\hat X}^2 \hat{X}_{\mu} \hat{X}^{\mu} 
- \frac{1}{2} \hat{\epsilon} \,\hat{X}_{\mu\nu} \hat{B}^{\mu\nu}  +{1 \over 2} m_{\hat{Z}}^2 \hat{Z}_\mu \hat{Z}^\mu \nl
&& + \ol{\chi} (i \slashed{D}  - m_\chi ) \chi
 -{\delta \over 4} (\ol{\chi^c} \chi + h.c.),
\label{eq:Lag}
\end{eqnarray}
where $\hat{X}_{\mu \nu} = \partial_{\mu} \hat{X}_{\nu} - \partial_{\nu} \hat{X}_{\mu}$
($\hat{B}_{\mu \nu} = \partial_{\mu} \hat{B}_{\nu} - \partial_{\nu} \hat{B}_{\mu}$) is the field strength tensor for the dark
photon $\hat{X}_\mu$ (for the $U(1)_Y$ gauge boson $\hat{B}_\mu$), 
$D_{\mu} = \partial_{\mu} + i g_X Q_\chi \hat{X}_{\mu}$ is the covariant derivative 
with $g_X$  the dark coupling constant, and we fix the dark charge of $\chi$, $Q_\chi =1$.
The above Lagrangian can be considered an effective theory of the UV-complete theory given, for example, in~\cite{Baek:2020owl}.
We assume the kinetic mixing between the dark photon and the $U(1)_Y$ gauge field is small, $\hat{\epsilon} \ll 1$.
We do not specify the origin of the dark gauge boson mass $m_{\hat{X}}$ and the $Z$-boson mass $m_{\hat{Z}}$.

The gauge fields can be written in terms of mass eigenstates: the photon $A_\mu$, the SM $Z$-boson, and the physical DP $Z'$.
As we will see in Section~\ref{sec:XENON1T}, we are interested in very small mixing parameter, $\hat{\ep} \sim 10^{-10}$. 
So it suffices to keep only the linear terms in $\hat{\ep}$ in the analysis.
In this case we get~\cite{Babu:1997st}
\bea
\hat{B}_\mu &=& c_W A_\mu - s_W Z_\mu -\hat{\ep} c_W^2  Z'_\mu, \nl
\hat{W}^3_\mu &=& s_W A_\mu + c_W Z_\mu - \hat{\ep} c_W s_W  Z'_\mu, \nl
\hat{X}_\mu &=&  Z'_\mu  + \hat{\ep} s_W  Z_\mu,
\label{eq:mixing}
\eea
where $c_W =\cos\theta_W (s_W =\sin\theta_W)$ with $\theta_W$  the Weinberg angle, and $\hat{W}^3_\mu$ is the neutral component of $SU(2)_L$ gauge boson.
In this approximation the gauge boson masses do not get corrections by the mixing (\ref{eq:mixing}): {\it i.e.}
$m_Z = m_{\hat{Z}}$ and $m_{Z'} = m_{\hat{X}}$.

The Dirac field $\chi$ splits into two Majorana mass eigenstates, $\chi_1$ and $\chi_2$, defined as~\cite{Baek:2020owl}
\begin{eqnarray}
\chi &  = & \frac{1}{\sqrt{2}} ( \chi_2 + i \chi_1 ) ,
\\
\chi^c &  = & \frac{1}{\sqrt{2}} ( \chi_2 - i \chi_1 ) ,
\\
\chi_1^c & = & \chi_1 , \ \ \  \chi_2^c = \chi_2, 
\end{eqnarray}
with masses
\begin{equation}
m_{\chi_1,\chi_2} = m_\chi \mp  y v_\phi \equiv m_\chi \mp \frac{1}{2} \delta .
\end{equation} 
We assume   $\delta \equiv m_{\chi_2} - m_{\chi_1}  > 0$. Then $\chi_1$ becomes the DM.    
The dark-gauge interactions of the DM and electron are~\cite{Baek:2020owl} 
\begin{eqnarray}
{\cal L} & \supset & 
- i g_X Z'_\mu \overline{\chi_2} \gamma^\mu \chi_1-\hat{\ep} e c_W Z'_\mu \overline{e} \gamma^\mu e. 
\end{eqnarray}
Note that the gauge interactions change the {\it flavour} of the dark fermions: $\chi_1 \leftrightarrow \chi_2$.
In the rest of the paper we fix $\delta = 2.8$ keV, $m_{Z'}=10^{-4}$ eV, and find $m_{\chi_1}$, $\alpha_X (=g_X^2/4\pi)$
which can explain the DM relic abundance, the small scale problem, and the XENON1T anomaly at the same time.

\section{The relic abundance}
\label{sec:relic}

In the early universe the dark sector (DS) can be in thermal equilibrium with the SM sector via process such as
%$\chi_i \chi_i \rightleftharpoons Z' Z'$ ($i=1,2$), 
$Z' f \rightleftharpoons \gamma(Z,g) f$, and $Z' \gamma(Z,g) \rightleftharpoons f \overline{f}$ with $f$ a SM fermion. 
In this case the Boltzmann equation for the DM number density reads~\cite{Griest:1990kh}
\bea
\frac{d n}{d t} + 3 H n = - \sum_{i,j=1,2} \langle \si_{ij} v \rangle (n_i n_j - n_i^{\rm eq} n_j^{\rm eq}),
\label{eq:Boltzmann}
\eea
where $n=n_1 + n_2$ with $n_i \equiv n_{\chi_i}$.
Following the procedure in~\cite{Griest:1990kh}, we obtain
the freeze-out temperature $T_f$ of the DM by solving
\bea
x_f = \log\frac{0.0382\, g_1  m_1 M_{\rm pl}  x_f^{1/2}  \langle \sigma_{\rm eff} v\rangle}{g_{*}^{1/2}},
\label{eq:x_f}
\eea
where $x_f \equiv m_1/T_f$ with $m_1 \equiv m_{\chi_1}$ and $M_{\rm pl} \simeq 1.22 \times 10^{19}$ GeV is the Planck mass.  The effective thermal-averaged cross section
$\langle \sigma_{\rm eff} v\rangle$ is obtained by
\bea
\langle \sigma_{\rm eff} v\rangle = \sum_{i,j=1,2} \langle \sigma_{ij} v\rangle r_i  r_j,
\quad r_i = \frac{g_i (1+\Delta_i)^{3/2} e^{-x_f \Delta_i}}{\sum_{i=1,2}g_i (1+\Delta_i)^{3/2} e^{-x_f \Delta_i}},
%\simeq {1 \over 4} (\langle \sigma_{11} v \rangle+\langle\sigma_{22} v\rangle) \simeq {1 \over 2}\langle \sigma_{11} v \rangle.
\eea
where $\Delta_i = (m_i -m_1)/m_1$.
Since $\Delta_2 (= \delta/m_1) \ll 1$, $x_f \Delta_2 \ll 1$, and $\sigma_{11} \simeq \sigma_{22} \gg \sigma_{12}$ in our scenario, we can approximate
\bea
\langle \sigma_{\rm eff} v\rangle \simeq {1 \over 2} \langle \sigma_{11} v\rangle.
\eea

Explicitly we get the dominant $s$-wave contributions to the DM annihilations to be
\begin{eqnarray}
\si_{ii} v&\simeq& \si v (\chi_i \chi_i \to Z' Z')   \simeq \frac{\pi \alpha_X^2}{m_1^2} +O(v^2), \nl
\si_{ij} v&\simeq& \si v (\chi_i \chi_j \to f \ol{f})  \simeq \frac{2 \pi \epsilon^2 N_c^f \al_{\rm em} \al_X (2 m_1^2 + m_f^2)(m_1^2-m_f^2)^{1/2}}{9 m_1^5}+O(v^2),
\label{eq:ann_x_section}
\end{eqnarray}
where $i \not= j$ ($i,j=1,2$), $N_c^f$ is the color factor of $f$, $\ep \equiv \hat{\ep} c_W$, and we used $m_{Z'}, \delta \ll m_1$.
We see that $\sigma_{ij} v$ ($i \not= j$) is $\epsilon^2$-suppressed and negligible compared to $\sigma_{ii} v$.

Let us comment on a possible issue in the annihilation cross section. At high energy, $s \to \infty$, the cross section, $\sigma(\chi_i \chi_i \to Z' Z')$, behaves like
\bea
\sigma(\chi_i \chi_i \to Z' Z') = \frac{\al_X^2 \pi \delta^2}{m_{Z'}^4} + O\left(1 \over s\right),
\eea
which violates the perturbative unitarity. So we need a UV completion of (\ref{eq:Lag}) to cure this problem.
For example, we can introduce Higgs field(s) coupled to $\chi_i$. In this UV completion the Higgs mediated contribution
to $\chi_i \chi_i \to Z' Z'$ is $p$-wave, and does not change the $s$-wave term in (\ref{eq:ann_x_section}). 
The resulting relic density obtained from $s$-wave contribution only from (\ref{eq:ann_x_section}) has corrections of order $O(3/x_f) \simeq O(0.1)$. 
So we can take (\ref{eq:Lag}) as a leading effective
Lagrangian for $\chi_i$ and $Z'$ for a large class of microscopic theories where other NP particles are integrated out.

By solving the Boltzmann equation (\ref{eq:Boltzmann}) with the condition (\ref{eq:x_f}) the final relic density from $s$-wave only in (\ref{eq:ann_x_section}) is obtained to be
\begin{align}
\Omega h^2  &\simeq \frac{2 \times 1.038 \times 10^{19} \, x_f \; {\rm GeV}^{-1} }{g_{*S}(T_f)/g_{*}^{1/2}(T_f) M_{\rm pl} \langle \sigma_{11} v \rangle} \nl
                      & \approx 0.12 \left(3.59 \times 10^{-3} \over \alpha_X \right)^2 \left(m_{\chi_1} \over 100 \, {\rm GeV}\right)^2,
\label{eq:Omega}
\end{align}
where $g_{*S}$ and $g_{*}$ are defined in~\cite{Kolb:1988aj}.

\section{The small scale problem}
\label{sec:small}

Given a DM with mass $m_1$, the value of the dark gauge coupling constant $\alpha_X$ to yield the correct relic abundance can be predicted
from (\ref{eq:Omega}). When these two parameters $m_1$ and $\al_X$
and the DP mass $m_{Z'}$ are known, 
we can calculate the elastic scattering, $\chi_1 \chi_1 \to \chi_1 \chi_1$, and inelastic scattering, $\chi_1 \chi_1 \to \chi_2 \chi_2$, cross sections for a given DM relative velocity $v$. Since for $m_{Z'} \ll m_1$ the non-perturbative Sommerfeld effect becomes
significant,  the perturbative
calculation cannot be applied here. Following the Ref.~\cite{Blennow:2016gde}, we solve the corresponding Schr\"odinger equation
in the CM-frame to calculate the two cross sections,
\begin{align}
\left[ -\frac{\nabla^2}{m_1}+V(\vec{r})\right] \Psi(\vec{r}) = \frac{k^2}{m_1} \Psi(\vec{r}),
\label{eq:Schroedinger}
\end{align}
where $\Psi$ is the $2 \times 1$ matrix wavefunction for the DM states with the upper component for the $\chi_1 \chi_1$ state and the lower component for the
$\chi_2 \chi_2$ state,  $\vec{r} = \vec{r}_1 -\vec{r}_2$ is the relative spatial coordinate of colliding DM particles, and $\vec{k}=m_1 \vec{v}/2$ is the relative momentum with $\vec{v}$ the
relative velocity.
The  potential  is written in the $2 \times 2$-matrix form,
\bea
V(r) =
\begin{pmatrix}
0 & -\frac{\alpha_X}{r} e^{-m_{Z'} r} \\
-\frac{\alpha_X}{r} e^{-m_{Z'} r} & 2 \delta\\
\end{pmatrix}.
\eea
As in~\cite{Blennow:2016gde}, we adopt the method suggested in~\cite{Ershov:2011zz} to solve the differential equation 
(\ref{eq:Schroedinger}) numerically. 
The system of  coupled radial equations in (\ref{eq:Schroedinger}) leads to numerical instability in a classically forbidden region.
The numerical stability is enhanced by using the modified variable phase method presented in~\cite{Ershov:2011zz}.
We introduce dimensionless parameters~\cite{Blennow:2016gde},
\bea
a = \frac{k}{\al_X m_1}, \quad
b = \frac{\al_X m_1}{m_{Z'}}, \quad
c=\sqrt{a^2 - {2 \de \over \al_X^2 m_1}},\quad
x=m_1 \al_X r,
\eea
in terms of which the radial part of the Schr\"odinger equation (\ref{eq:Schroedinger}) becomes
\bea
\Bigg[ {d^2 \over dx^2} - \frac{\ell (\ell+1)}{x^2} + 
\begin{pmatrix}
a^2 & 0 \\
0 & c^2
\end{pmatrix} \Bigg] \chi(x)
=
\begin{pmatrix}
0 & -{1 \over x} e^{-x/b} \\
-{1 \over x} e^{-x/b} & 0
\end{pmatrix} \chi(x),
\label{eq:chi}
\eea
where $\chi(x) = x R(x)$ with $\Psi(\vec{r}) = R(r) Y_{\ell m}(\theta,\phi)$. As in~\cite{Ershov:2011zz} we write the solution in the form of $2 \times 2$ matrix
\bea
\chi_{ij}(x) = \big( f(p_i x) \delta_{ik} - h^{(+)}(p_i x) M_{ik}(x) \big) \alpha_{kj}(x),
\eea
where $i,j,k=1,2$, the repeated indices are to be summed but not for the free indices $i,j$.
We use $f(x)=x j_\ell(x), h^{(+)}(x)= i x h^{(1)}_\ell(x)$ which are the solutions when the potential in (\ref{eq:chi}) is set to be zero and $a=c=1$.
Here $j_\ell(x)$ ($h_\ell^{(1)}(x)$) are the spherical Bessel functions (the spherical Hankel functions of the first kind).
The boundary condition $\chi_{ij}(x=0)=0$ leads to $M_{ij}(x=0)=0$.
We can write the scattering amplitude in terms of the matrix $M(x=\infty)$, or equivalently, in terms of a unitariy matrix $S_\ell \equiv 1 - 2 i M^\ell(x=\infty)$ as
\bea
f(\theta) = -i \sum_{\ell=0}^\infty (2 \ell+1) P_\ell(\cos\theta) \begin{pmatrix} {1 \over 2 k} & 0 \\ 0 & {1 \over 2 k'}\end{pmatrix} (S_\ell-1),
\eea
where $k' =\sqrt{k^2 -2 m_1 \delta}$. 

The differential cross sections for the scattering of identical particles are obtained by
\bea
\left(\frac{d \sigma}{d \Omega}\right)_\xi = \left| f(\theta) + \xi f(\pi-\theta)\right|^2,
\eea
where $\xi=+1 (-1)$ if the spatial wave function is symmetric (antisymmetric) under particle exchange.
Assuming the DM is unpolarized, we average over the spin states to get
\bea
\frac{d \sigma}{d \Omega} ={1 \over 4} \left(\frac{d \sigma}{d \Omega}\right)_{\xi=+1}+ {3 \over 4} \left(\frac{d \sigma}{d \Omega}\right)_{\xi=-1},
\eea
where the first (the second) term is the contribution from the spin singlet (triplet) state with symmetric (antisymmetric) spatial wave funtion.
Then the inelastic scattering cross section is obtained by
\bea
\sigma_{\rm inel}(\chi_1 \chi_1 \to \chi_2 \chi_2) &=& 
 {k' \over 2k} \int \left[{1 \over 4}\left| f_{21}(\theta) +f_{21}(\pi-\theta)\right|^2 
+{3 \over 4} \left| f_{21}(\theta) -f_{21}(\pi-\theta)\right|^2 \right] d\Omega \nl
&=& \frac{4 \pi }{k k'} \sum_{\ell=0}^\infty \zeta_\ell (2\ell+1) \left| M^\ell_{21}(x=\infty)\right|^2,
\label{eq:sigma_inel}
\eea
where $\zeta_\ell=1/2\,  (3/2)$ for $\ell=$even (odd) and
the values of $M_{21}^\ell$'s are to be evaluated at $x=\infty$.
For the elastic scattering cross section which is used to solve the small scale problem, we consider the viscosity~\cite{Blennow:2016gde} and the momentum-transfer~~\cite{Kahlhoefer:2017umn}  cross section.
The viscosity cross section is given by
\bea
&&\sigma_{\rm el}^V(\chi_1 \chi_1 \to \chi_1 \chi_1) \nl
&=& {1 \over 2}\int 
\left[{1 \over 4} \left| f_{11}(\theta)+f_{11}(\pi-\theta)\right|^2 + {3 \over 4} \left| f_{11}(\theta)-f_{11}(\pi-\theta)\right|^2 \right]
\sin^2\theta d\Omega \nl
&=& \frac{4 \pi}{k^2} \sum_{\ell=0}^\infty \zeta_\ell\Bigg[ \left| M_{11}^\ell \right|^2 \frac{2(2\ell+1)(\ell^2+\ell-1)}{(2\ell-1)(2\ell+3)} 
-(M_{11}^\ell (M_{11}^{\ell+2})^* +c.c.) \frac{(\ell+1)(\ell+2)}{2\ell+3} \Bigg]. 
\label{eq:sigma_el_V}
\eea
For the momentum transfer cross section it is more convenient to evaluate the spin singlet and triplet contributions separately,
\bea
\sigma_{\rm el}^T(\chi_1 \chi_1 \to \chi_1 \chi_1) &=& {1 \over 4}\sigma_{\rm el}^{T, {\rm Singlet}}+ {3 \over 4}\sigma_{\rm el}^{T, {\rm Triptlet}}, 
\eea
with
\bea
\sigma_{\rm el}^{T,{\rm Singlet}}
&=& {1 \over 2}\int 
 \left| f_{11}(\theta)+f_{11}(\pi-\theta)\right|^2 
(1-|\cos\theta|) d\Omega \nl
&=& \frac{8 \pi}{k^2} \sum_{\ell,\ell'={\rm even}} (2\ell'+1) M_{11}^\ell {M_{11}^{\ell'}}^*\left[
\delta_{\ell \ell'}-\left((\ell+1) f_{\ell',\ell+1} +\ell f_{\ell',\ell-1}\right)
\right], \nl
\sigma_{\rm el}^{T,{\rm Triplet}}
&=& {1 \over 2}\int 
 \left| f_{11}(\theta)-f_{11}(\pi-\theta)\right|^2 
(1-|\cos\theta|) d\Omega \nl
&=& \frac{8 \pi}{k^2} \sum_{\ell,\ell'={\rm odd}} (2\ell'+1)M_{11}^\ell {M_{11}^{\ell'}}^*\left[
\delta_{\ell \ell'}-\left((\ell+1) f_{\ell+1,\ell'} +\ell f_{\ell-1,\ell'}\right)
\right],
\label{eq:sigma_el_T}
\eea
where
\bea
f_{\ell,\ell'}=\frac{(-1)^{(\ell+\ell'+1)/2} \ell! \ell'!}{2^{\ell+\ell'-1}(\ell-\ell')(\ell+\ell'+1)
\left[\left(\ell \over 2\right)!\right]^2 \left[\left(\ell'-1 \over 2\right)!\right]^2}, \quad \text{with
$\ell=$even, $\ell'=$odd}.
\eea

To obtain $M^\ell(x=\infty)$ we first transform $M$-matrix to $U$-matrix~\cite{Ershov:2011zz},
\bea
U_{ij}(x) = f(p_i) h^{(+)}(p_i) \delta_{ij} - h^{(+)}(p_i) M_{ij}(x) h^{(+)}(p_j),
\eea
which gives numerically more stable solutions.
From (\ref{eq:chi}) we get a coupled first-order differential equation for $U(x)$,
\bea
U'_{ij}(x) &=& p_i \delta_{ij} + p_i {g'(p_i x)  \over g(p_i x)} U_{ij}(x) + U_{ij}(x) {g'(p_j x)  \over g(p_j x)} p_j -U_{ik}(x) {1 \over p_k} \hat{V}_{kl} U_{lj}(x),
\label{eq:U_eq}
\eea
where
\bea
\hat{V}(x) =
\begin{pmatrix}
0 & -{1 \over x} e^{-x/b} \\
-{1 \over x} e^{-x/b} & 0
\end{pmatrix}.
\eea
The $S$-matrix is related to the $U$-matrix as
\bea
S_{ij} = \frac{h^{(-)}(p_i x)}{h^{(+)}(p_i x)} \delta_{ij} + 2 i \frac{1}{h^{(+)}(p_i x)} U_{ij}(x) \frac{1}{h^{(+)}(p_j x)},
\label{eq:S}
\eea
where $h^{(-)}(x)=-i x h^{(2)}_\ell(x)$, $h^{(2)}_\ell(x)$ the spherical Hankel function of the second kind, and we take $x\to \infty$.
We solved (\ref{eq:U_eq}) numerically with initial condition
\bea
U_{ij}(x_0) = f(p_i x_0) h^{(+)}(p_i x_0) \delta_{ij} \simeq \frac{p_i x_0}{2\ell+1} \delta_{ij},
\eea
where we take $x_0=0.01$. The results are not very sensitive to the value of $x_0$ as long as $x_0 \ll 1$.
When we integrate (\ref{eq:U_eq}) to large $x$, $x=x_\infty (\gg 1)$, we take reasonably large value of $x_\infty$ in such a way that not only 
the $S$-matrix obtained in (\ref{eq:S}) keeps unitarity (when $k,k'>0$) but also $U(x_\infty)$ converges to a constant value.

\begin{table}[t]
\renewcommand{\arraystretch}{1.5}
%\scriptsize
\begin{center} 
\begin{tabular}{|c||c|c|c|}\hline
                                             & BP1          &   BP2             & BP3        \\ 
\hline\hline
$m_{\chi_1} \; (\rm GeV)$         &  $70$             &   $100$  & $120$\\
\hline 
$\alpha_X$                & $2.51 \times 10^{-3}$ & $3.59 \times 10^{-3}$ & $4.31 \times 10^{-3}$\\
\hline
$\sigma_{\rm inel} \; (\rm pb)$  & $3.93 \times 10^{14} $ & $1.78 \times 10^{14}$ & $5.71 \times 10^{13}$\\
\hline
 $\sigma^V_{\rm el}/m_{\chi_1} \; (\rm cm^2/g)$ & $10.6 (1.95, 1.86 \times 10^{-3})$ 
   &  $3.19(0.871,1.38\times 10^{-3})$ 
   & $1.49 (0.270, 8.12 \times 10^{-4})$\\
\hline
$\sigma^T_{\rm el}/m_{\chi_1} \; (\rm cm^2/g)$ & $7.63 (1.22, 1.12 \times 10^{-3})$ 
    & $2.14(0.645,5.88\times 10^{-4})$ 
    & $0.962 (0.191, 1.74 \times 10^{-4})$\\
\hline
$\epsilon$   & $5.39 \times 10^{-10}$ & $1.14 \times 10^{-9}$ & $2.02 \times 10^{-9}$\\
\hline
%%%
\end{tabular}
\end{center}
\caption{
The results for the three benchmark masses $m_\chi=70, 100, 120$ GeV. 
In the 2nd line the required dark gauge coupling constant $\al_X$ to give the correct relic abundance, $\Omega_{\rm DM} h^2 =0.119$, is given.
In the 3rd--5th lines, the predictions of the cross sections for the inelastic scattering, $\sigma_{\rm inel}(\chi_1 \chi_1 \to \chi_2 \chi_2)$, 
and elastic scattering, $\sigma_{\rm el}^{V(T)}(\chi_1 \chi_1 \to \chi_1 \chi_1)/m_{\chi_1}$, are listed. We have fixed the DM velocity $v\approx220 \, {\rm km/s}$ in the Milky Way for
the inelastic scattering calculation which is relevant for the XENON1T experiment. 
For the elastic cross sections, we used the DM velocity $v \approx 30 (220, 3000)\, {\rm km/s}$ which corresponds to the dwarf galaxies (the Milky
Way, the bullet cluster).
In the last line we show the value of kinetic mixing parameter ($ \epsilon =\hat{\epsilon} c_W$) which explains the XENON1T excess.
We fixed the other parameters: $\delta=2.8\,{\rm keV}$, $m_{Z'}=10^{-4}\,{\rm eV}$.
}
\label{tab:benchmark}
% \end{tiny}
\end{table}

The predictions for the elastic and inelastic DM annihilation cross sections for benchmark DM masses $m_{\chi_1} = 70, 100, 120$ (GeV) are shown in Table~\ref{tab:benchmark} along with other results.
To calculate the cross sections we have fixed, $v=30, 220, 3000 \, {\rm km/s}$, corresponding to a typical DM velocity at the dwarf galaxies, the Milky Way, and the
clusters of galaxies, respectively.
For the inelastic scattering we show the results only with $v=220\, {\rm km/s}$ which is relevant for the XENON1T experiment.
We can see that the inelastic cross sections have the correct values to solve the small scale problems.
The momentum-transfer cross sections $\si_{\rm el}^T$ are smaller than the viscosity cross sections $\si_{\rm el}^V$
as can be expected from the suppression of $1-|\cos\theta|$ compared to $\sin^2\theta$. But they are similar in size and either of
them can be used to measure the effect of the elastic scattering.
The elastic cross sections are also highly velocity-dependent and can evade the constraints from
the Milky Way and the galaxy clusters such as the bullet cluster.
We note that the results in Table~\ref{tab:benchmark} are not very sensitive to $m_{Z'}$ as long as $m_{Z'} \ll \delta$.

Some comments are in order. For $v=30, 220 \,{\rm km/s}$, we sum only up to $l \lesssim  30$ in (\ref{eq:sigma_inel}), 
(\ref{eq:sigma_el_V}) and (\ref{eq:sigma_el_T}) because $M_{ij}^\ell \to 0$ rapidly beyond 
$\ell \sim  30$. 
For $v=3000 \,{\rm km/s}$ the solution does not decrease easily as $\ell$ increases 
and the unitarity of $S$-matrix begins to be violated by ${\cal O}(1)$ when $\ell \gtrsim 60$, and we stop near $\ell\sim 60$
to keep the unitarity. So the cross sections for $v=3000 \,{\rm km/s}$ in Table~\ref{tab:benchmark} are expected to have ${\cal O}(1)$ errors.
To make sure the elastic cross sections are suppressed to satisfy the constraints for this DM velocity we 
cross-checked the elastic cross sections using the Born approximation which is a good approximation in this regime. 
The elastic scattering occurs at the second order of the Born expansion.
The result, $\sigma^V_{\rm el}/m_{\chi_1}=1.37 \times 10^{-3}, 2.26 \times 10^{-3}, 1.88 \times 10^{-3} \,{\rm cm^2/g}$(for $m_{\chi_1}=70, 100, 120 \,{\rm GeV}$, respectively),
indeed shows that they are small enough to satisfy the constraint $\sigma_{\rm el}/m_{\chi_1} \lesssim 0.5 \,{\rm cm^2/g}$.

\section{The XENON1T excess}
\label{sec:XENON1T}

Now the excited state $\chi_2$ produced by the inelastic scattering decays promptly back into $\chi_1$ and $Z'$, $\chi_2 \to \chi_1 Z'$, with 100\% branching ratio.
Since the mass difference $\delta =m_{\chi_2} - m_\chi$ is fixed to be $2.8 \, {\rm keV}$, the energy of $Z'$ is also fixed to be that of the $\delta$.
The relativistic $Z'$ is absorbed in a xenon atom in the XENON1T experiment and ejects an electron with energy close to $2.8\, {\rm keV}$ via a photoelectric-like effect.

The flux of $Z'$ per unit energy within the solid angle $\Delta \Omega$, coming
from the inelastic scattering $\chi_1 \chi_1 \to \chi_2 \chi_2$ and the subsequent decay $\chi_2 \to \chi_1 Z'$, is obtained by
\begin{align}
\frac{d \Phi_{Z'}}{d E_{Z'}} =\Delta \Omega \, \frac{r_\odot}{8\pi} \, \frac{d N_{Z'}}{d E_{Z'}} \, \left(\frac{\rho_\odot}{m_{\chi_1}} \right)^2 \,  \langle \sigma_{\rm inel} v\rangle \, \bar{J},
\end{align}
where $r_\odot \simeq 8.33$ kpc is the distance from the Earth to the galactic center (GC), $\rho_\odot \simeq 0.3\, {\rm GeV/cm^3}$ is the local DM density,
the $J$-factor, $\bar{J} = \int \frac{d \Omega}{\Delta \Omega} \int_{l.o.s.} \frac{ds}{r_\odot} \left(\rho(r) \over \rho_\odot \right)^2$,
%$r=(s^2+r_\odot^2-2 s r_\odot \cos\theta)^{1/2}$, 
is the line-of-sight integration
of the DM density squared, and the energy spectrum is given by $\frac{d N_{Z'}}{d E_{Z'}} = 2 \delta(E_{Z'}-\delta)$ in our model.
Considering the $Z'$-flux from the full sky, {\it i.e.} $\Delta \Omega=4 \pi$, we get $\bar{J} \simeq 2.20$ by using the cored isothermal DM density profile~\cite{Jimenez:2002vy,Ng:2013xha}:
\begin{align}
\rho(r) &= \rho_\odot \left[\frac{r}{r_\odot}\right]^{-\gamma} \left[\frac{1+(r_\odot/r_s)^\alpha}{1+(r/r_s)^\alpha}\right]^{\frac{\beta-\gamma}{\alpha}},
\end{align}
where $\{\alpha,\beta,\gamma,r_s\} = \{2,2,0,3.5\,{\rm kpc}\}$\footnote{We obtain $\bar{J}=2.97$ for the NFW profile with $\{\alpha,\beta,\gamma,r_s\} = \{1,3,1,20\,{\rm kpc}\}$.}.
Then the differential event rate of the dark-photoelectric effect per ton of the xenon target per year can be written as~\cite{Chiang:2020hgb}
\begin{align}
\frac{d R}{d E_e} &=\int d E_{Z'} \, \frac{d \Phi_{Z'} }{d E_{Z'} } \, \frac{\sigma_{Z'}(E_{Z'})}{m_{Xe}} \frac{1}{\sqrt{2\pi} \sigma} e^{-\frac{(E_e-E_{Z'})^2}{2\sigma^2}} \epsilon(E_e),
\end{align}
where $E_e$ is the emitted electron energy, $m_{Xe}$ is the mass of a xenon atom,
and $\sigma_{Z'}(E_{Z'}) =\epsilon^2 \sigma_\gamma(E_{Z'})$ is the dark-photoelectric cross section of the xenon atom at the energy $E_{Z'}$.
The Gaussian function simulates the smearing effect of the electron energy by the detector resolution with~\cite{Aprile:2020tmw}
\begin{align}
\frac{\sigma}{E_e} = \frac{a}{\sqrt{E_e/{\rm keV}}} + b,
\end{align}
where $a=0.3171 \pm 0.0065$ and $b = 0.0015 \pm 0.0002$.
The function $\epsilon(E_e)$ is the total detector efficiency reported in~\cite{Aprile:2020tmw}.
We obtain $\sigma_\gamma(2.8\, {\rm keV}) \simeq 2.0 \times 10^5\, {\rm barn}$~\cite{XCOM}.
To fit the XENON1T data we find the required values of $\ep$ values are $\ep=5.39 \times 10^{-10}, 1.14 \times 10^{-9}, 2.02 \times 10^{-9}$  for $m_{\chi_1} =70, 100, 120$ GeV, respectively.

\begin{figure}[htbp]
\begin{center}
\includegraphics[width=0.6\textwidth]{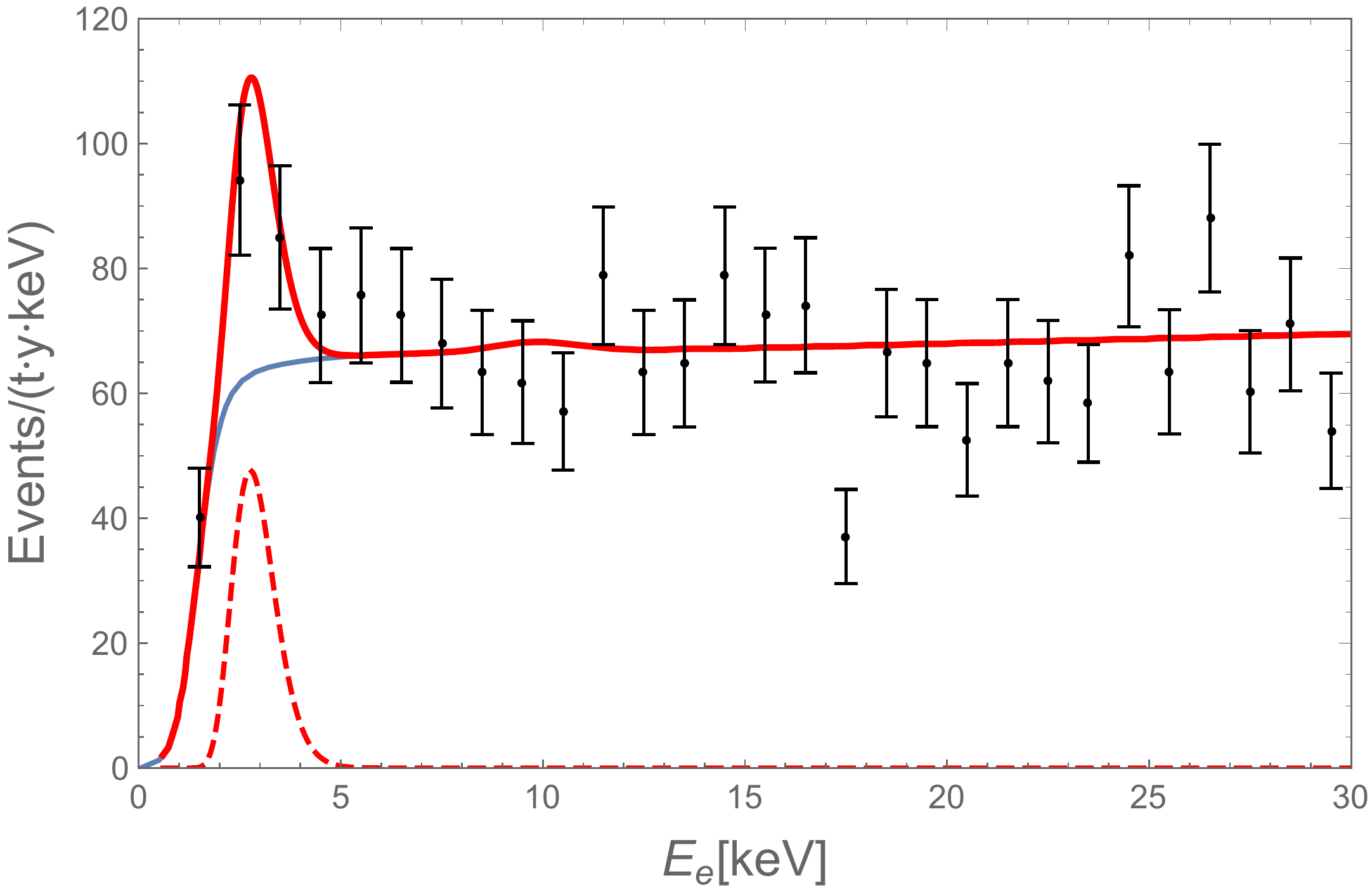}
\caption{The differential event rate (solid red curve) for a benchmark point, $m_{\chi_1}=100\, {\rm GeV}$, 
$\alpha_X = 3.59 \times 10^{-3}$,
$\epsilon=1.14 \times 10^{-9}$, $\delta=2.8\,{\rm keV}$, and $m_{Z'}=10^{-4}\,{\rm eV}$. 
The dashed curve is the contribution of the NP signal only.
The experimental data and the background blue curve
are extracted from~\cite{Aprile:2020tmw}.
}
\label{fig:XENON1T}
\end{center}
\end{figure}

In Fig.~\ref{fig:XENON1T} we show the resulting differential event rate (solid red curve) for
a benchmark point, $m_{\chi_1}=100\, {\rm GeV}$, 
$\alpha_X = 3.59 \times 10^{-3}$,
$\epsilon=1.14 \times 10^{-9}$, $\delta=2.8\,{\rm keV}$, and $m_{Z'}=10^{-4}\,{\rm eV}$. 
The dashed curve is the contribution of the NP signal only.
The experimental data and the background blue curve
are extracted from~\cite{Aprile:2020tmw}.

These values of $\ep$ are consistent with the current experimental constraints~\cite{Jaeckel:2013ija,An:2013yua}.
The null observation of the dark matter at the DM-nucleon scattering experiments can be explained by the $\ep^2$-suppressed cross section of the relevant 
scattering $\chi_1 q \to \chi_2 q$ and also by the inelasticity of the scattering.

\section{Conclusions}
\label{sec:conclusion}

Although the existence of dark matter is well-established, its particle nature is almost unknown.
The small scale problem and the recent observation of the excess in electron-recoil at XENON1T experiment may reveal the nature of dark matter.
We studied a dark matter model which can address these issues while explaining its abundance in the universe.

In the model the Majorana dark matter candidate $\chi_1$ has its excited partner $\chi_2$ with mass difference $\delta\simeq2.8$ keV.
The dark mass is about $100$ GeV and it can explain the current relic density by the thermal freeze-out mechanism whose main annihilation
process is $\chi_1 \chi_1 (\chi_2 \chi_2) \to Z' Z'$. We find that the necessary dark gauge coupling is $\alpha_X \sim 10^{-3}$.

The light $Z'$ can also mediate (in)elastic scattering $\chi_1 \chi_1 \to \chi_{1(2)} \chi_{1(2)}$. We solve the Schr\"odinger equation numerically to calculate
the cross sections. We get the elastic cross section large enough to explain the small scale problems $\sigma_{\rm el}/m_{\chi_1} \sim 1 \; {\rm cm^2/g}$ . 

The dark sector can communicate with the SM sector through kinetic mixing parameterized by $\ep \sim 10^{-10}$.
In the current universe the rate for the up-scattering $\chi_1 \chi_1 \to \chi_2 \chi_2$ followed by $\chi_2 \to \chi_1 Z'$ can be enhanced by the small
$Z'$ mass $m_{Z'} \ll \delta$. The energetic $Z'$ is absorbed by the xenon atom at the XENON1T detector via the mechanism similar to the photoelectric effect.
The above mentioned values of $\ep$ and $\delta$ can explain the spectrum and the excess event rate observed by the XENON1T.

\acknowledgments
This work was supported in part by the National Research Foundation of Korea(NRF) grant funded by
     the Korea government(MSIT), Grant No. NRF-2018R1A2A3075605.

%\bibliographystyle{JHEP}
%\bibliography{Inelastic_Z2}

\providecommand{\href}[2]{#2}\begingroup\raggedright\endgroup

\end{document}